\documentclass[manuscript]{acmart}
\usepackage{svg}
\AtBeginDocument{%
  }

\setcopyright{none}
\renewcommand\footnotetextcopyrightpermission[1]{} 
\copyrightyear{2025}
\acmYear{2025}
\setcopyright{rightsretained}
\acmConference[CI 2025]{Collective Intelligence Conference}{August 4--6, 2025}{San Diego, CA, USA}
\acmBooktitle{Collective Intelligence Conference (CI 2025), August 4--6, 2025, San Diego, CA, USA}
\acmDOI{10.1145/3715928.3758069}
\acmISBN{979-8-4007-1489-4/2025/08}

\begin{document}

%%
%% The "title" command has an optional parameter,
%% allowing the author to define a "short title" to be used in page headers.
\title{Seeing Twice: How Side-by-Side T2I Comparison Changes Auditing Strategies}

%%
%% The "author" command and its associated commands are used to define
%% the authors and their affiliations.
%% Of note is the shared affiliation of the first two authors, and the
%% "authornote" and "authornotemark" commands
%% used to denote shared contribution to the research.
\author{Matheus Kunzler Maldaner}
\email{mkunzlermaldaner@ufl.edu}
\orcid{0009-0007-6299-5831}
\affiliation{%
  \institution{University of Florida}
  \city{Gainesville}
  \state{Florida}
  \country{USA}
}

\author{Wesley Hanwen Deng}
%should we put emails for everyone?
%\email{larst@affiliation.org}
\affiliation{%
  \institution{Carnegie Mellon University}
  \city{Pittsburgh}
  \state{Pennsylvania}
  \country{USA}}

\author{Jason I. Hong}
%\authornotemark[1]
\affiliation{%
  \institution{Carnegie Mellon University}
  \city{Pittsburgh}
  \state{Pennsylvania}
  \country{USA}}

\author{Ken Holstein}
%\authornotemark[1]
\affiliation{%
 \institution{Carnegie Mellon University}
 \city{Pittsburgh}
 \state{Pennsylvania}
 \country{USA}}

\author{Motahhare Eslami}
%\authornotemark[1]
\affiliation{%
  \institution{Carnegie Mellon University}
  \city{Pittsburgh}
  \state{Pennsylvania}
  \country{USA}}

%%
%% By default, the full list of authors will be used in the page
%% headers. Often, this list is too long, and will overlap
%% other information printed in the page headers. This command allows
%% the author to define a more concise list
%% of authors' names for this purpose.
\renewcommand{\shortauthors}{Kunzler Maldaner al.}

%%
%% The abstract is a short summary of the work to be presented in the
%% article.
\begin{abstract}
    While generative AI systems have gained popularity in diverse applications, their potential to produce harmful outputs limits their trustworthiness and utility. A small but growing line of research has explored tools and processes to better engage non-AI expert users in auditing generative AI systems. In this work, we present the design and evaluation of MIRAGE, a web-based tool exploring a ``contrast-first'' workflow that allows users to pick up to four different text-to-image (T2I) models, view their images side-by-side, and provide feedback on model performance on a single screen. In our user study with fifteen participants, we  used four predefined models for consistency, with only a single model initially being shown. We found that most participants shifted from analyzing individual images to general model output patterns once the side-by-side step appeared with all four models; several participants coined persistent “model personalities’’ (e.g., \textit{cartoonish}, \textit{saturated}) that helped them form expectations about how each model would behave on future prompts. Bilingual participants also surfaced a language-fidelity gap, as English prompts produced more accurate images than Portuguese or Chinese, an issue often overlooked when dealing with a single model. These findings suggest that simple comparative interfaces can accelerate bias discovery and reshape how people think about generative models.
\end{abstract}

%%
%% The code below is generated by the tool at http://dl.acm.org/ccs.cfm.
%% Please copy and paste the code instead of the example below.
%%


%%
%% Keywords. The author(s) should pick words that accurately describe
%% the work being presented. Separate the keywords with commas.

%% A "teaser" image appears between the author and affiliation
%% information and the body of the document, and typically spans the
%% page.
\begin{teaserfigure}
  \centering
  \includegraphics[width=1\textwidth]{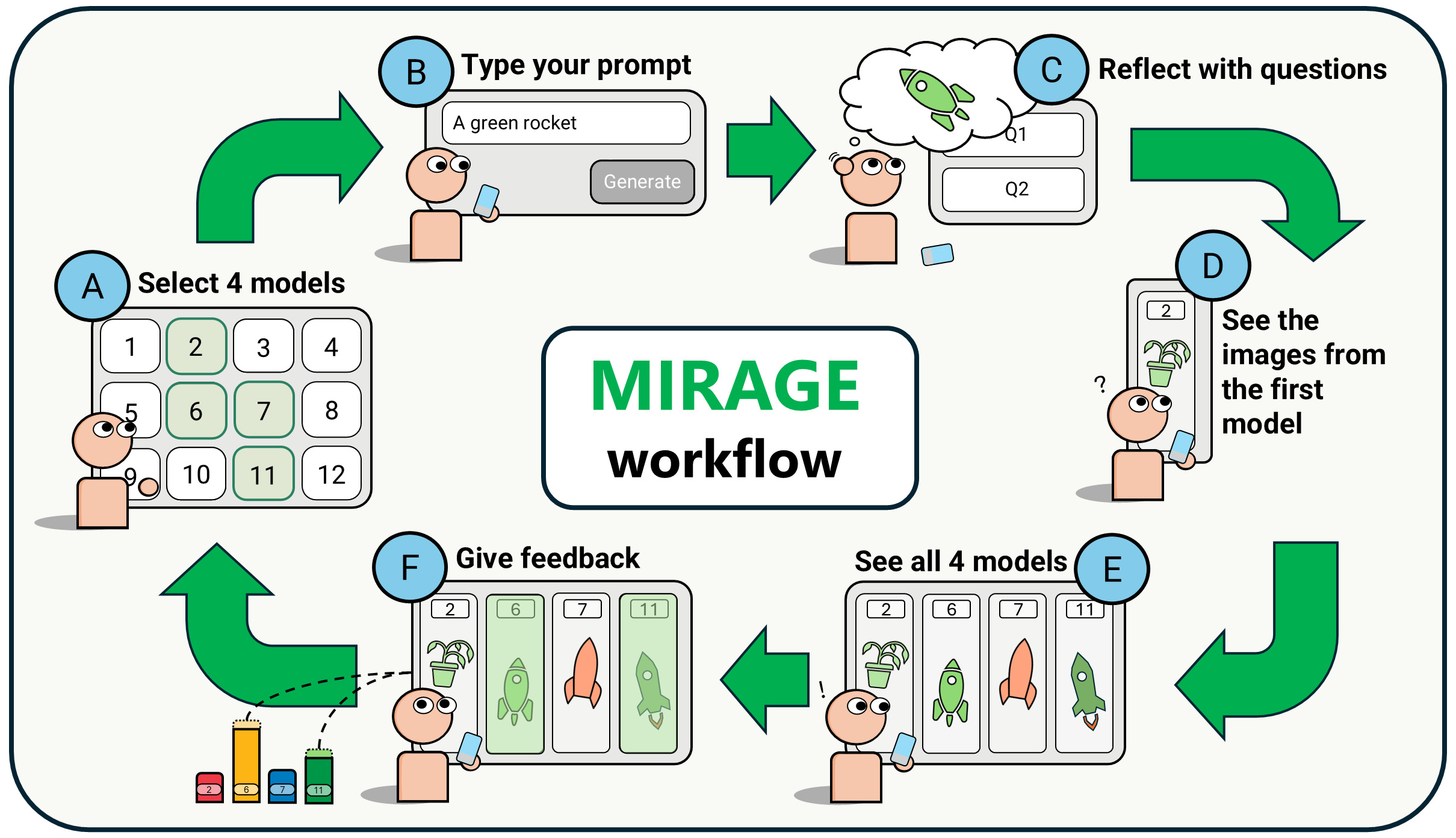}
  \caption{MIRAGE supports the comparison of different text-to-image models displayed side-by-side in a single window. Users can select up to four models to be displayed (A) and enter a prompt of their choice (B). While the generation happens in the background, users are asked to reflect on what they expect to see (C, Q1) and why they chose that specific prompt (C, Q2). Users in this study first saw the outputs of just one of the four models (D) and then of all of them (E). Finally, the users can analyze all of the images and select which models matched their original expectations while pointing out any details they might have not noticed before (F).}
  \label{fig:mirage-workflow}
  \Description{The MIRAGE workflow interface showing steps for comparing outputs of text-to-image models. The figure includes labeled sections: (A) selecting up to four models, (B) entering a text prompt, (C) answering reflective questions about user expectations, (D) viewing outputs from a single model, and (E) viewing outputs from all selected models side-by-side. The final step allows users to analyze and compare the images for alignment with their expectations.}
\end{teaserfigure}

% \received{20 February 2007}
% \received[revised]{12 March 2009}
% \received[accepted]{5 June 2009}

%%
%% This command processes the author and affiliation and title
%% information and builds the first part of the formatted document.
\maketitle

\section{Introduction}

% Despite growing popularity, generative AI such as text-to-image (T2I) systems are known to produce harmful outputs, such as images that reinforce societal stereotypes or that contribute to political misinformation~\cite{bianchi2023easilyaccessiblet2i}. For example, a recent news article suggested that Stable Diffusion, a popular open-source T2I model, rarely depicted women as doctors, lawyers, or judges, and often implied that men with dark skin commit crimes, reinforcing harmful gender and racial biases \cite{Nicoletti2023humans, buolamwini2018gender}. \looseness=-1

Auditing is crucial to ensure fairness in AI systems~\cite{Sandvig2014AuditingA}. At a high level, AI auditing refers to a process of repeatedly testing an algorithm with inputs and observing the corresponding outputs, in order to understand its behavior and potential external impacts~\cite{metaxa2021auditing,birhane2024ai}. Recognizing the power of diverse end users in surfacing harmful behaviors in AI systems that might otherwise be overlooked by small, homogeneous groups of AI developers, recent research has explored the potential of engaging users in auditing AI systems~\cite{shen2021everydayalgorithmauditing,devos2022toward,deng2023understanding}. Some platforms, such as Hugging Face and OpenAI, have developed interfaces using side-by-side comparisons for GenAI outputs, as a method to help users or red-teamers review and audit GenAI outputs \cite{redteaming100}. However, there remains a lack of scientific understanding of how side-by-side comparison can support activities such as AI auditing and red-teaming, as well as other potential human-AI interactions. \looseness=-1

% Existing theories, such as Variation Theory for human concept learning \cite{ling2012variation}, suggest that contrasting outputs can help users identify critical features and develop a more nuanced understanding. 

In this paper, we explore whether \textbf{viewing outputs from multiple text-to-image models side-by-side can help users identify details they might miss when only looking at a single model's output.} We extend our previous work on (1) the design and development of MIRAGE\footnote{MIRAGE stands for \textbf{M}ulti-model \textbf{I}nterface for \textbf{R}eviewing and \textbf{A}uditing \textbf{Ge}nerative Text-to-Image AI}\cite{maldaner}, a web application supporting multi-model side-by-side comparison while soliciting users' perception changes, (2) an expansion of the findings from a study with fifteen users, and (3) the exploration of future work to leverage the MIRAGE framework to further explore the effects of side-by-side comparison for AI auditing and beyond.

\section{Related Work}

A small but growing line of work has explored developing tools for engaging AI developers and users in testing and auditing AI models. Current examples include side-by-side comparators such as IndieLabel, LLM-Comparator, and WeAudit \cite{Lam2022endusersaudits,kahng2024llmcomparatorvisualanalytics,deng2025weaudit}, and industry dashboards from Hugging Face, OpenAI, Replicate, and Chatbot Arena for red-teaming \cite{huggingface-judge,openai-compare,Replicate,chatbot-arena,redteaming100}. This paper extends prior work by developing a web-based interface to explore how reviewing outputs from multiple \emph{distinct} T2I models simultaneously affects users' abilities and strategies of conducting AI audits.

\section{System Overview}

% MIRAGE is a web interface for auditing text-to-image models via side-by-side comparison. Users enter a prompt (Fig.~\ref{fig:mirage-interface}A), pick up to four models with “Show Models,” and answer short reflection questions (B). The first model to finish appears alone (C); once all four are ready, users compare 32 images (8 per model) side-by-side (D). Answers never influence generation—they simply populate the audit report.

MIRAGE provides a simple workflow where users select up to four models from the thirty available models, enter a prompt, and see the outputs side-by-side. Each model generates eight images, for a total of thirty-two images across all models. For this study, the same four models were fixed across sessions. The new MIRAGE extension codebase has been made modular, enabling new models to be added with minimal effort.

To mask latency, MIRAGE employs “micro-interaction” buffering~\cite{bernstein2010soylent}. While participants answer two reflection questions, images render in the background, so most results appear instantly when participants look back~\cite{Nah2004}. Several participants even assumed the images were pre-cached rather than generated live. The implementation, deployment, and interface details appear in Appendix~\ref{appendix:implementation} and Appendix~\ref{appendix:interface}. Please visit \url{mirage.weaudit.org} to access MIRAGE \footnote{We plan to live demo MIRAGE during the conference}.

\section{Pilot User Study}

\begin{figure}[b]
\centering
\includegraphics[width=0.75\columnwidth]{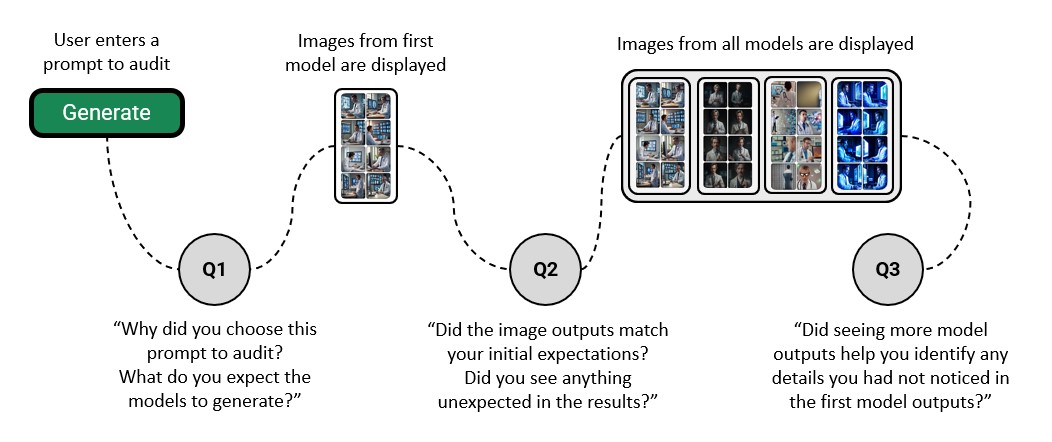} 
\caption{MIRAGE User Study Workflow}
\Description{MIRAGE User Study Workflow}
\label{fig:workflow}
\end{figure}

Fifteen volunteers completed a 45-minute, IRB-approved session and were compensated with a \$15 gift card. Each session started with a predefined prompt (``a fancy dinner party''), and users then contributed two self-chosen prompts based on their lived experiences. While MIRAGE worked in the background, users answered two questions (``Why did you choose this prompt?” and “What do you expect to see?''), giving the system time to generate unseen outputs from four fixed T2I models: SDXL-Lightning, Kandinsky 2.2, Stable Diffusion, and Latent Consistency. To reduce latency bias, SDXL-Lightning's outputs were always displayed first because of its generation speed. After rating that single-model view, participants saw the full four-panel grid (Fig. \ref{fig:workflow}), noted any newly observed details, and described how their auditing strategy shifted. Six sessions were conducted in Portuguese or Chinese, providing an early look at multilingual prompts. Full demographic details appear in Appendix~\ref{appendix:demographics}.

\section{Preliminary Findings}

Overall, participants reported that MIRAGE was generally clear and intuitive to use. Some also appreciated the question flow (see Figure \ref{fig:workflow}), which helped them reflect on their own biases and articulate the reasoning behind their prompt choices during auditing. In the sections below, we expand on two key insights into how MIRAGE inspired users to develop new auditing strategies both in terms of low-level model output details and high-level model ``personalities,'' expanding the affordances offered by prior user-AI auditing tools
\cite{Lam2022endusersaudits, deng2025weaudit}.

%The preliminary user study of MIRAGE provided several insights into its usefulness and usability. 

%To start with, all fifteen participants found MIRAGE easy to use, with a clear and intuitive workflow. P3 especially appreciated the question flow (see Figure \ref{fig:workflow}), which allowed them to reflect on their own biases and articulate the reasons behind their prompt choices for auditing. Bilingual participants also observed that identical prompts written in English produced noticeably higher-fidelity images than when entered in Portuguese or Chinese, underscoring gaps in current multilingual support. These results are broken down in the sections that follow.

%Although participants were initially instructed to test only two prompts of their choice, many became highly engaged and chose to enter more prompts on their own. For example, P14 tested five prompts, including variations of ``A fancy dinner party," because she was curious about how different models interpreted her ideas. This heightened engagement highlights the system's potential to spark curiosity and broaden perspectives.

\subsection{Auditing Strategies Inspired by Contrasting Multi Model Outputs} \label{subsec:auditing-strategies}

%% Can probably remove the part below, seems redundant. Also look through Miro board / documents for more participants that share these feelings.
The exposure to multiple images led participants to develop new auditing strategies. In particular, P1, P4, P5, and P13 reported that they focused on inspecting individual images when there was only one model but focused on reviewing the overall image output distribution when there were multiple models. When asked how seeing more model outputs affected their strategies for analyzing the images, participants noted a clear shift in their approach. 
% Several participants (P4, P5, P13) highlighted a transition from focusing on individual image details when analyzing a single model to examining broader trends and distributions across multiple models. 
P4 specifically mentioned that with multiple models, they were \textit{``less likely to look at individual images and more at surface-level outputs and trends,"} emphasizing a more holistic approach. 
% Similarly, P5 explained that multi-model comparisons allowed them to analyze groups of images against one another, enabling the identification of shared characteristics and differences between models, while P13 noted that having multiple models allowed for comparing the outputs across models rather than just aligning them with personal expectations.
% \textit{``If there is only one model, I would compare what I have in mind with what the model has, but when there are four models, I can compare their outputs against one another.''} Some participants also referenced concepts of intra- versus inter-model variance, with P5 and P13 pointing out how this comparison framework provided a deeper understanding of each model's unique behavior. 
P11 and P14 described how multi-model viewing helped them notice details they had missed in single-model outputs, such as biases in age, color schemes, or stylistic consistencies, with P11 remarking that they ``had not noticed how good the first model was until they saw the others.'' 

% Towards the end of the study, participants like P6 and P10 emphasized how this approach helped them rapidly identify the strengths and weaknesses of specific models, such as Stable Diffusion's inability to generate faces, or how certain models excelled or failed in producing consistent styles. P8 noted that comparing multiple models fixed expectations for future analyses, as they began to associate specific characteristics with particular models, which shaped subsequent evaluations.

In addition, viewing outputs from additional models enabled participants to identify new image details that could lead to potentially harmful biases that were not recognized when viewing the outputs of a single model. For instance, when P1 saw the first model's output for the prompt ``An intern working in the city of Pittsburgh,'' they only noted that all workers were depicted outdoors. However, after seeing additional model outputs, they observed that darker-skinned workers were depicted wearing construction clothes, while lighter-skinned workers were depicted in office attire, carrying keyboards and notebooks. In another example, participants who initially expected the prompt ``A fancy dinner party'' to generate images of white males noticed that all tables were long and rectangular, a more Western style, as opposed to the round tables commonly found in Asian cultures.

% P14 reflected that, ``Seeing all models allowed me to open my mind and see different perspectives from my preconceived expectations, since the same sentence can mean different things.'' Similarly, P15 reflected on the surprising outcomes from the generated images, stating that ``[although] these images did not match my original expectations, [they] opened my mind to what a different version of a fancy dinner could look like.''

\subsection{Exploring Model Personalities} \label{subsec:personalities}
In addition to assisting participants in auditing AI-generated images, we also found that side-by-side comparison helped participants compare and understand model output styles, or the model ``personalities.'' 
%(...)
%In particular, when asked how they perceived each model, participants reported distinct ``personalities'' for each system. 
For example, several participants (P5, P7, P8, P11, P10) praised ByteDance Sdxl as offering the “best quality” and “most variation,” though P4 found its outputs “least culturally diverse.” Latent Consistency elicited comments like “cartoonish” (P5) or “exaggerated” (P3), and some (P4, P6, P7) noted a lack of diversity in the outputs. Interestingly, P7 said that its consistency was beneficial for generating coherent storylines. Kandinsky was described as “colorful and vibrant” (P8, P5, P10). Finally, participants found Stable Diffusion’s proportions “off” (P3, P5, P6, P8, P11, P10), but P7 felt it generally “matched expectations.” These varied impressions underscore that each model may cater to different user needs, prompting future inquiry into how style, realism, or diversity impacts user preference.

\section{Future Work}

% In this paper, we introduced and utilized MIRAGE, a web-based tool designed to facilitate side-by-side comparisons of multiple AI text-to-image models. Our pilot user study, involving fifteen participants, showed that MIRAGE was successful in helping users spot new details and gain new perspectives by comparing multiple models side-by-side that were not apparent when viewing the outputs of a single model. In addition, participants developed new auditing strategies by examining trends across multiple models, and they recognized distinct ``personalities'' in model outputs, providing deeper insights into model behavior and user preferences.
Looking ahead, we propose the following future use cases with the goal of bridging the gap between everyday users and developers.

\textbf{(1) Controlled Experiments}:
Run large-scale controlled experiments to \emph{quantify} how side-by-side workflows improve bias detection and how factors such as identity or AI literacy mediate those gains.  

% %\dhw{I still feel like anonymous auditing can be combined with the model leader board section}
% \textbf{(2) Anonymous Auditing}; Companies and developers are often reluctant to give public access to their AI models due to the risk of exposing proprietary technology to competitors or facing public backlash if the models produce harmful content \cite{deng2023understanding}. This reluctance limits the ability of companies to gather valuable feedback from everyday users, who are social actors engaged in the daily use of algorithm auditing systems \cite{shen2021everydayalgorithmauditing}. In line with a functionality provided by Chatbot Arena \cite{chiang2024chatbotarenaopenplatform}, we envision adapting MIRAGE to act as a bridge between everyday users and developers of proprietary text-to-image models. In this future application, developers can submit their models to MIRAGE, which will anonymize and deliver them to everyday users who can leverage their lived experiences and social backgrounds to provide feedback and locate harmful behaviors. 

\textbf{(2) Anonymous Auditing}: Add an \emph{anonymous-auditing} mode so developers can submit proprietary models that are anonymized in MIRAGE, letting diverse users surface harms without exposing trade secrets.

\textbf{(3) Model Leaderboard}: Launch a community \emph{leaderboard} that ranks T2I systems by real-user preference clicks, creating a feedback loop that motivates developers to fix issues highlighted through MIRAGE.

Through collaboration between users from varying backgrounds and AI developers, MIRAGE has the potential to drive meaningful advancements in AI transparency and auditing.

% Check if this is correct Wesley:
% \section{Acknowledgments}
% \noindent This work was supported by the National Science Foundation (NSF) program on Fairness in AI in collaboration with Amazon under Award No. IIS-2040942.

\bibliographystyle{ACM-Reference-Format}
\bibliography{references}

\appendix

\section{Appendix}

\subsection{Participant Demographics}
\label{appendix:demographics}

The participants in our pilot study encompassed a group of fifteen individuals aged 18 to 76, spanning ethnicities, educational backgrounds, and genders. Participants were recruited through convenience sampling, and the group included native speakers of English, Portuguese, and Chinese, which enabled the exploration of cross-linguistic and cultural differences in text-to-image (T2I) model auditing. Participants' experience using artificial intelligence and T2I models and tools varied widely, with proficiency levels ranging from minimal exposure (1) to advanced expertise (5) on a 1–5 scale. This diversity allowed us to evaluate the usability and effectiveness of MIRAGE across a broad spectrum of users, incorporating varied cultural contexts, technical expertise, and lived experiences, as summarized in Table \ref{tab:participants}.

\begin{table*}[t]
\centering
\begin{tabular}{|c|c|c|c|c|c|c|c|}
\hline
Participant & Age & Gender & Ethnicity & Educational Level & Language & AI Exp. & T2I Exp. \\ \hline
P01 & 20 & Female & Asian & Bachelor's degree & English & 3 & 3 \\ \hline
P02 & 21 & Male & Indian & Bachelor's degree & English & 5 & 3 \\ \hline
P03 & 19 & Female & Black & Bachelor's degree & English & 3 & 1 \\ \hline
P04 & 22 & Female & Asian & Master's degree & English & 2 & 1 \\ \hline
P05 & 18 & Female & Latino, White & High School & English & 3 & 2 \\ \hline
P06 & 22 & Male & Latino, White & Bachelor's degree & Portuguese & 3 & 2 \\ \hline
P07 & 76 & Male & Latino, White & Master's degree & Portuguese & 1 & 1 \\ \hline
P08 & 63 & Male & Latino, White & Bachelor’s degree & Portuguese & 5 & 5 \\ \hline
P09 & 25 & Male & Asian &  Bachelor’s degree  & Chinese, English & 2 & 1 \\ \hline
P10 & 20 & Female & Asian & Bachelor's degree & English & 3 & 4 \\ \hline
P11 & 25 & Female & Asian & Master’s degree & English & 3 & 3 \\ \hline
P12 & 45 & Female & Asian & High School & Chinese, English & 3 & 1 \\ \hline
P13 & 20 & Male & Asian & Bachelor's degree & English & 4 & 3 \\ \hline
P14 & 50 & Female & Latino, White & Bachelor’s degree & Portuguese & 1 & 1 \\ \hline
P15 & 21 & Female & White & Bachelor's degree & English & 5 & 4 \\ \hline
\end{tabular}
\caption{Demographics and AI experience of participants in the MIRAGE pilot study.}
\label{tab:participants}
\end{table*}

\subsection{Technical Implementation}
\label{appendix:implementation}

\begin{figure}[H]
\centering
\includegraphics[width=0.8\columnwidth]{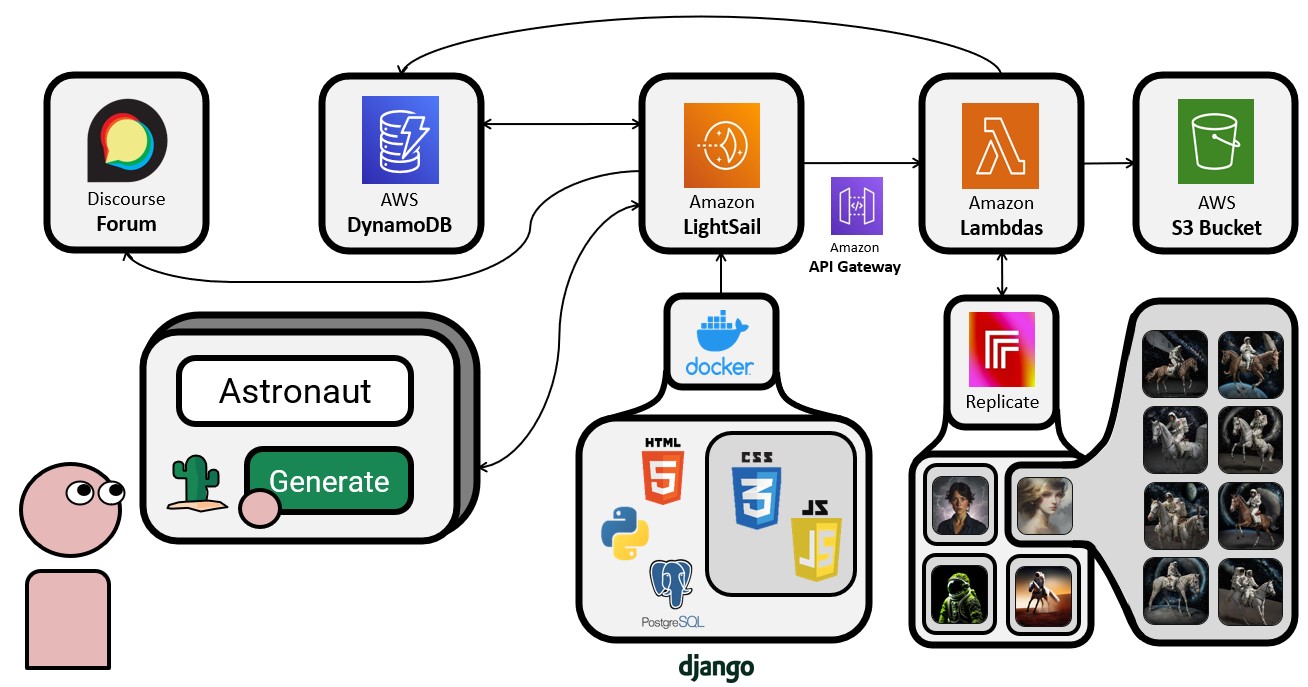}
\caption{MIRAGE Technical Implementation.}
\label{fig:backend}
\end{figure}

In order to achieve our goal, we first developed the MIRAGE web application (Figure \ref{fig:backend}), and hosted it on Amazon Lightsail with a Docker container packaging a Django project. Image generation for specific models is handled through API calls to Replicate, a centralized service that works as a hub for open-source models. Replicate stores the generated images in their servers for up to an hour, so we save all images to our own Amazon S3 Bucket with unique IDs for persistence. The images and their associated audits are referenced in an Amazon DynamoDB table for easy retrieval and analysis. Backend computations, including calls to Replicate, are performed using AWS Lambda functions, accessed through Amazon API Gateway. Although not used in our study, the Discourse API integration is designed to facilitate community discussions about audit findings in future iterations of MIRAGE. Lastly, the design of the system is modular, allowing new T2I models to be added as they become available. While users are encouraged to try the system deployed at \url{mirage.weaudit.org}, we acknowledge that research teams may want to explore different research questions using MIRAGE. We therefore plan to open-source the application in the coming months.

%%
%% End of file `sample-sigconf.tex'.
\subsection{Application Interface}
\label{appendix:interface}
\begin{figure}[H]
    \centering
    \includegraphics[width=\linewidth]{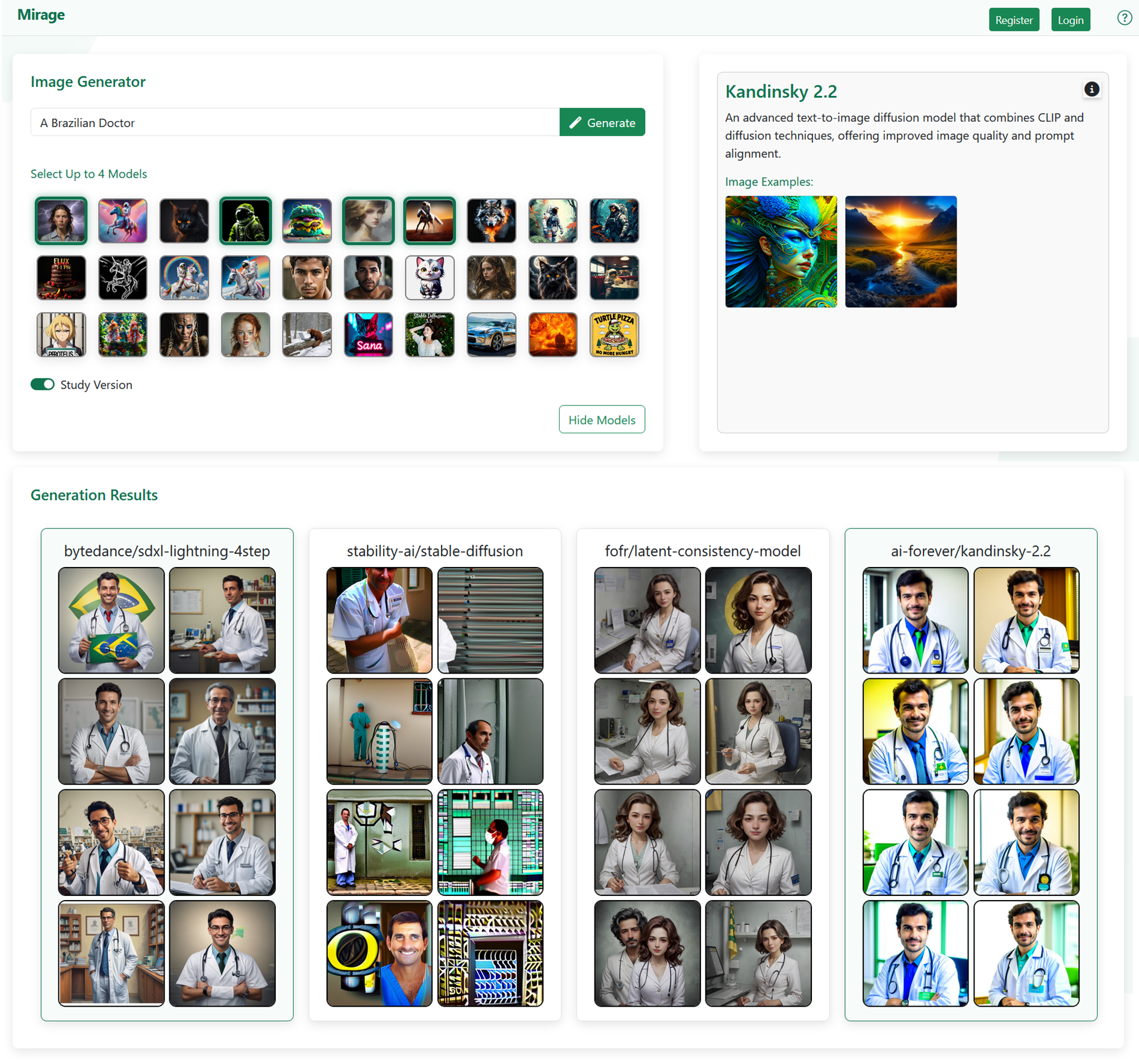}
    \caption{MIRAGE Main Page. Image Generator (upper-left) lets users type a prompt, toggle the “Study Version” option for the user study scenario, and pick up to four models from a total of thirty thumbnail icons. Hovering over a model opens the Model Preview panel (upper-right), which shows example images. The upper-right panel is also populated with reflection questions during the study workflow. Pressing Generate fills the Generation Results area with eight images per model, arranged in four parallel columns, so users can compare thirty-two outputs at a glance.}
    \label{fig:interface3}
\end{figure}

\begin{figure}[H]
    \centering
    \includegraphics[width=\linewidth]{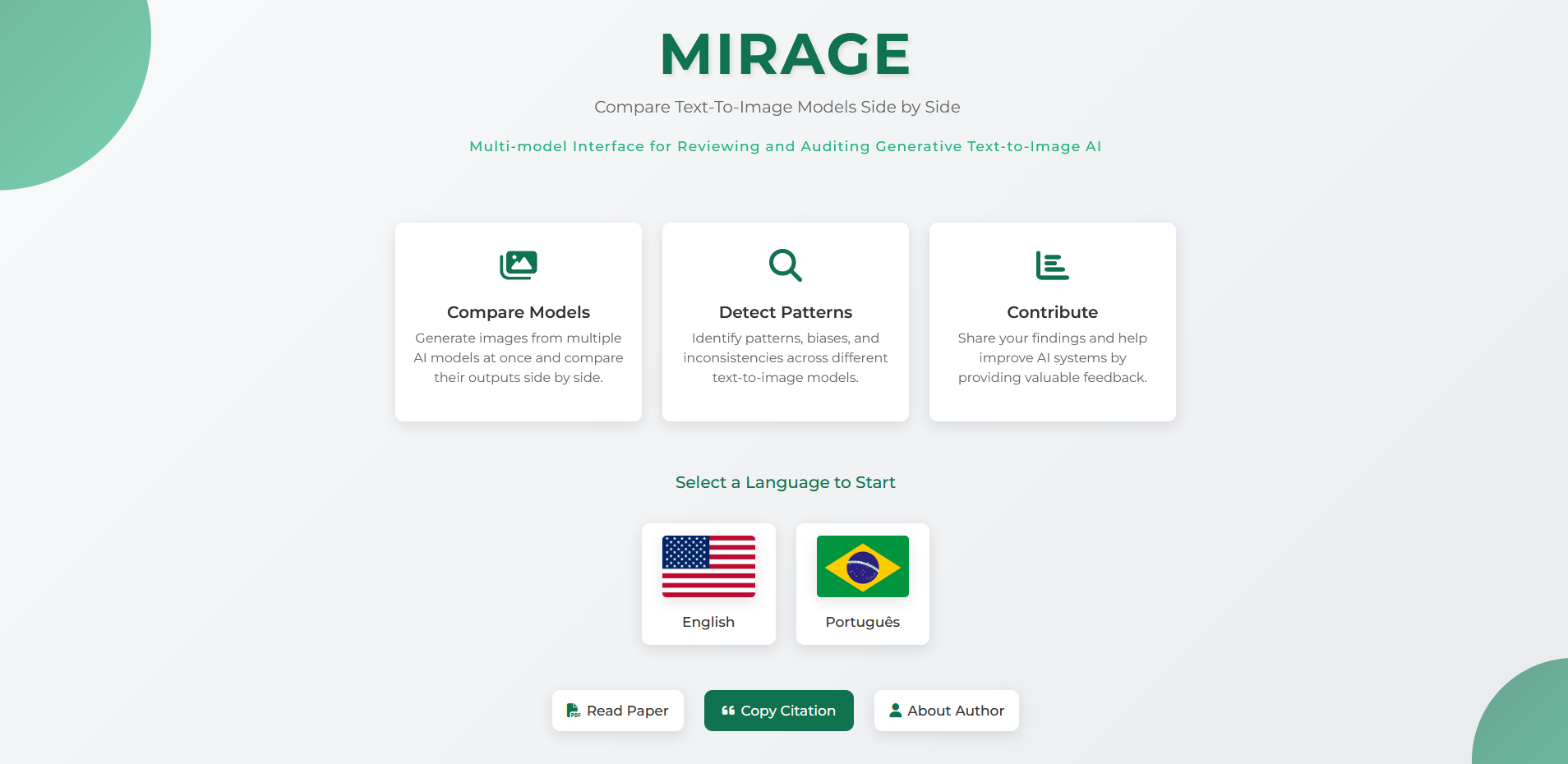}
    \caption{MIRAGE Landing Page. Users can read the purpose of the tool, select a language and read previously published work.}
    \label{fig:interface1}
\end{figure}

\begin{figure}[H]
    \centering
    \includegraphics[width=\linewidth]{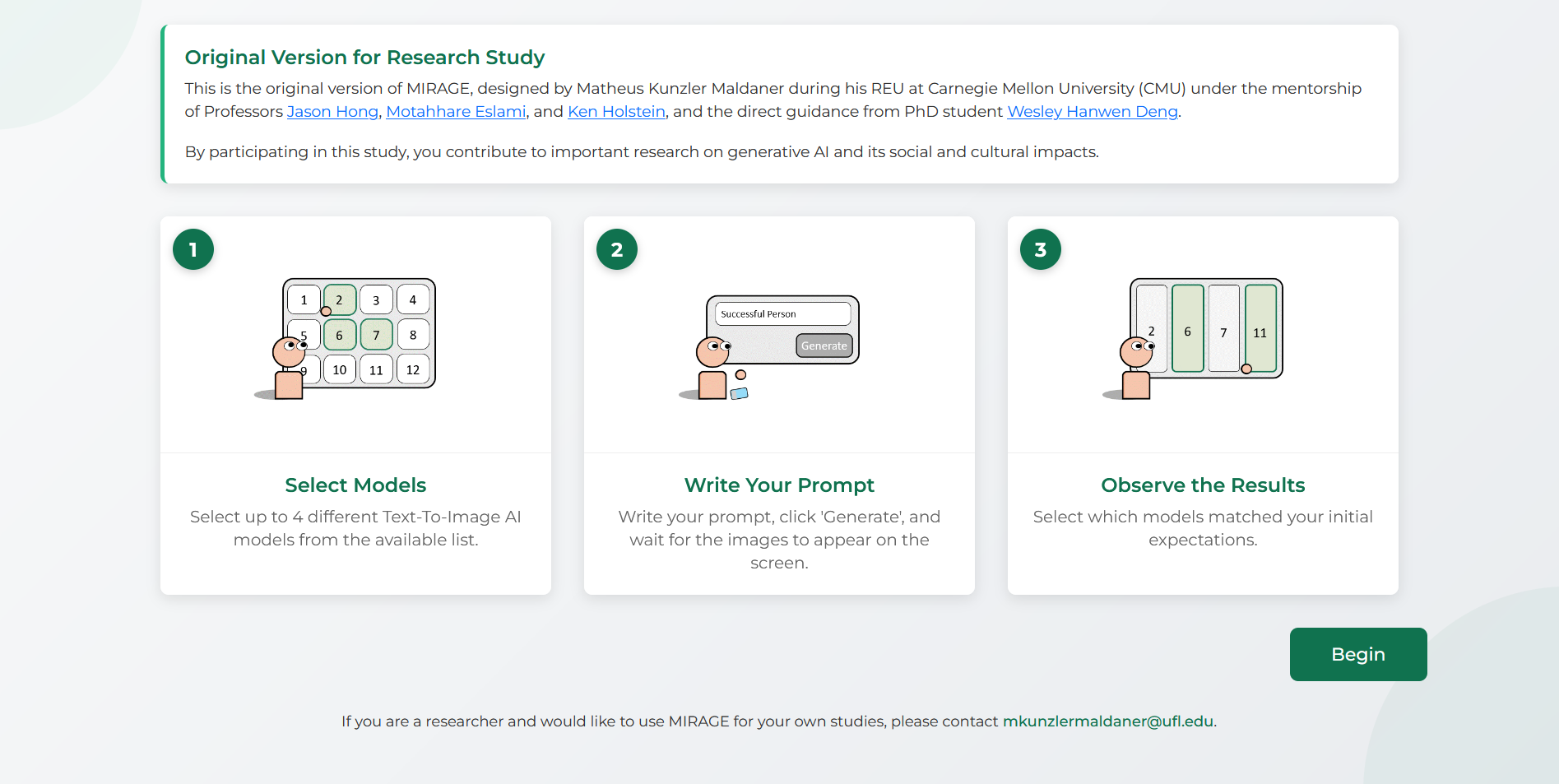}
    \caption{MIRAGE Instructions Page. Users are shown a simple three-step approach to using the tool.}
    \label{fig:interface2}
\end{figure}

In order to reduce friction, we sought to create a minimalist and straightforward application. All users described MIRAGE as very easy to use. Figure~\ref{fig:interface3} shows the overall interface of our tool, broken down into three clearly labeled main areas. As seen in Figure~\ref{fig:interface2}, we are open to tailoring MIRAGE for specific research studies. Figure~\ref{fig:interface1} shows a ``Portuguese'' option for a customized version of MIRAGE designed specifically for a Brazilian research group.

\end{document}